\documentclass[preprint,showpacs,preprintnumbers,amsmath,amssymb]{revtex4}

\begin{document}
\title{Quantum Anti-Piracy Storage of Classical Data}
\author{Guang-Ping He}
\affiliation{Department of Physics and Advanced Research Center,
Zhongshan University, Guangzhou 510275, China}

\begin{abstract}
A scheme is proposed which stores classical data in 4-state
quantum registers. It can achieve the following goal: the
classical data can always be read unambiguously, while the quantum
registers cannot be copied. Therefore the data provider can always
distinguish the original quantum registers from piracy copies.
Examples of application are also given.
\end{abstract}

\maketitle

\newpage

\section{Introduction}

Piracy has become a worldwide problem of the whole society in recent years.
It has such a serious impact on the computer software and entertainment
industry, as well as literature and economy etc., that even our ordinary
life is involved. Laws were established and actions were taken against
piracy. But on the technical aspect, though countless anti-piracy methods
have been developed throughout the years, there were always new piracy
techniques coming up. Why did this happen? It is because there is no
theoretical limit on copying data within classical cryptography. In some
sense piracy is inevitable, because as long as the data can be unambiguously
read (which is a must for the user), it can be copied. For example, as long
as a film stored in a DVD can be decoded into image signals for playback, it
can be recorded by a camera or a VCR. We cannot hope that these re-recording
behaviors can be avoided technically, unless we do not want the film to be
visible to human eyes at all. The best we can expect is that the media that
carrying these data should not be duplicable. That is, if the images were
encoded again and stored in another DVD, we hope that this piracy copy
should be distinguishable from the original one. But in classical
cryptography, as there is no limit on data cloning, any classical
information (manufacture codes, labelling, stickers, etc.) of the media can
be copied perfectly in principle. Hence the security of all existing
classical anti-piracy methods has to rely on certain computational
assumptions, e.g. the hardness of factoring and the existence of
``trapdoor'' functions. It is proven \cite{Shor} that if quantum computer
becomes reality in the future, all these computational assumptions can be
broken easily, putting an end to all classical anti-piracy methods.

On the other hand, the rise of quantum cryptography \cite{Wiesner,BB84}
manifested that with quantum methods, a higher security level can be
expected. The security of these new cryptographic protocols is guaranteed by
the basic principles of quantum mechanics alone, thus surpasses its
classical counterpart. Therefore it is natural to ask whether quantum
cryptography can also be applied to anti-piracy purposes. In this paper,
basing on the quantum no-cloning theorem \cite{No-cloning}, a new and simple
approach is proposed which stores the classical data inside quantum states.
The important feature of the approach is: the user can always read the
classical data unambiguously, while no one but the original provider can
recreate the quantum states that carrying these data. Consequently, even
though a malevolent user can create some other quantum states storing the
same classical data, these states is distinguishable from the original ones.
This makes the approach useful for anti-piracy.

\section{The Scheme}

For conciseness, here we will focus on the idea setting without
the errors caused by technical problems. But in fact, with proper
modification (e.g. quantum error correcting codes), the scheme can
be adjusted to fit realistic settings. Let $\left| \alpha
_{0}\right\rangle $, $\left| \beta _{0}\right\rangle $, $\left|
\alpha _{1}\right\rangle $\ and $\left| \beta _{1}\right\rangle $
denote the orthogonal states of a 4-state quantum
register. When a provider wants to store a classical $n$-bit string $%
c=c_{1}c_{2}...c_{n}$, he simply performs the following {\bf Storing Protocol%
}:

For $i=1$ to $n$, the provider randomly chooses $\theta _{i,c_{i}}\in
\lbrack 0,2\pi )$ and prepares a quantum register in the state $\left| \psi
_{i}\right\rangle =\cos \theta _{i,c_{i}}\left| \alpha _{c_{i}}\right\rangle
+\sin \theta _{i,c_{i}}\left| \beta _{c_{i}}\right\rangle $. Then he gives
all these $n$ quantum registers to the user, while keeping all $\theta
_{i,c_{i}}$ secret.

\bigskip

When the user receives the registers, he runs the following {\bf Reading
Protocol} to read the classical data $c$:

For each $i$, the user measures the register $\left| \psi _{i}\right\rangle $%
\ with the projection operator $P=\left| \alpha _{0}\right\rangle
\left\langle \alpha _{0}\right| +\left| \beta _{0}\right\rangle \left\langle
\beta _{0}\right| $. He knows that $c_{i}=0$\ ($c_{i}=1$) if the projection
is successful (failed).

\bigskip

Now suppose that there are $n$ quantum registers $\left| \psi _{i}^{\prime
}\right\rangle $ ($i=1,...,n$) which store $c$. When the provider want to
check whether they are the original registers prepared by himself, he can
run the following {\bf Checking Protocol}:

For $i=1$ to $n$, the provider tries to project $\left| \psi _{i}^{\prime
}\right\rangle $ into the state $\left| \psi _{i}\right\rangle =\cos \theta
_{i,c_{i}}\left| \alpha _{c_{i}}\right\rangle +\sin \theta _{i,c_{i}}\left|
\beta _{c_{i}}\right\rangle $. He concludes that the registers are original
if all the projections are successful.

\bigskip

The security of this protocol is straight followed from the quantum
no-cloning theorem \cite{No-cloning}. Since $\theta _{i,c_{i}}$ is kept
secret from the user and can be varied within the range $[0,2\pi )$, the
register $\left| \psi _{i}\right\rangle $\ can never be cloned perfectly. If
a malevolent user wants to fake the register with probabilistic cloning or
even by guess, the fake register $\left| \psi _{i}^{\prime }\right\rangle $
cannot be projected into $\left| \psi _{i}\right\rangle $ without a
non-vanished error rate $\varepsilon $. Therefore faking $n$ registers can
only pass the Checking Protocol with the probability $(1-\varepsilon )^{n}$,
which is exponentially small as the length of the string $c$ increases.

Thus it can be seen that this protocol achieves the following goal: Any user
can perfectly retrieve the classical data stored in the quantum registers
without the help of the provider. Meanwhile, the provider can check whether
the quantum registers are original. That is, though a cheater can create
many sets of quantum registers storing the same classical data, the provider
can always distinguish them from the original.

Note that after the reading, the state of $\left| \psi _{i}\right\rangle $\
is perfectly undisturbed so that it can be read again at any time.
Meanwhile, as long as the registers remain original, the checking process
will not affect the quantum states at all. Thus they can be checked again
and again, without affecting the readability. Also, the checking can be
performed not only by the provider himself, but also any authorized person
who gets $\theta _{i,c_{i}}$\ from him.

\section{Applications}

Now we give an example on how the protocol can be used for anti-piracy
purposes. Up to now, the codes of the computer softwares are usually stored
on classical media, e.g. CD-Rom. A malevolent user can perfectly copy them
and sells them to others along with the serial numbers (which can also be
perfectly copied). In fact, any anti-piracy marks (such as the printing on
the CD, or some hidden codes inside the software) can also be copied.
Therefore all these copies look exactly the same to the original provider.
They cannot be distinguished even through on-line registration, which is an
anti-piracy method very commonly used today. And if the provider put limits
on the on-line registration, e.g. by limiting the times or frequency of the
registration, it may potentially infringer the right of legal users. But
with the new approach proposed above, this problem can be solved. The
software manufacturer can store the software codes in quantum registers with
our Storing Protocol. Obviously the user can easily read the code with the
Reading Protocol. When on-line registration is needed, the manufacturer can
ask the user to return some of the quantum registers, and run the Checking
Protocol to see whether they are original. From the discussion above we can
see that piracy copies of the software will inevitably be distinguished,
while only the original one can register on-line successfully. Thus improved
security is achieved.

Our protocol can also realize an interesting kind of authentication. Suppose
that a commander sends a messenger to deliver a message $c$ to a general.
They do not mind the messenger knowing $c$, but the general needs to ensure
that the content of $c$ is indeed from the commander and has not been
altered by the messenger or anyone else. This goal can be achieved as
follows. The commander and the general share beforehand a set of $\theta
_{i,0},\theta _{i,1}\in \lbrack 0,2\pi )$ ($i=1,...,n$, $\theta _{i,0}\neq
\theta _{i,1}$). At a later time, the commander encodes $c$ using the
Storing Protocol, and sends the messenger to deliver the quantum registers
to the general. The general first decodes $c$ with the Reading Protocol,
then with his knowledge on $c$ and $\theta _{i,c_{i}}$, he can check whether
$c$ is original with the Checking Protocol. This process is valid since no
one but the commander and the general knows $\theta _{i,\bar{c}_{i}}$.
Though it is possible for a malevolent messenger (or other cheaters) to
shift a quantum register from the state $\left| \psi _{i}\right\rangle =\cos
\theta _{i,c_{i}}\left| \alpha _{c_{i}}\right\rangle +\sin \theta
_{i,c_{i}}\left| \beta _{c_{i}}\right\rangle $ to $\left| \psi _{i}^{\prime
}\right\rangle =\cos \theta _{i,c_{i}}\left| \alpha _{\bar{c}%
_{i}}\right\rangle +\sin \theta _{i,c_{i}}\left| \beta _{\bar{c}%
_{i}}\right\rangle $ with unitary transformations, the result is not the
correct state encoding $\bar{c}_{i}$. Being ignorant of $\theta _{i,\bar{c}%
_{i}}$, it is impossible for the cheater to fake the correct state $\cos
\theta _{i,\bar{c}_{i}}\left| \alpha _{\bar{c}_{i}}\right\rangle +\sin
\theta _{i,\bar{c}_{i}}\left| \beta _{\bar{c}_{i}}\right\rangle $\ without a
non-vanished error rate $\varepsilon $. That is, even if the cheater
possesses all the quantum registers encoding $c$, he cannot change the bit $%
c_{i}$ into $\bar{c}_{i}$ flawlessly. As a result, altering many bits of $c$
can be successful only with an exponentially small probability. Thus we see
that all $\theta _{i,c_{i}}$\ together acts as a quantum signature which
comes inherently with the classical data, keeping the data secure from being
altered or faked.

Note that our protocol is difference from the conjugate coding in quantum
money \cite{Wiesner}. The conjugate coding also used quantum registers to
store some information of the provider, which can be checked by himself or
anyone authorized by him. However, the quantum registers contains no
readable information for the user. The classical data (e.g., the face value
of the quantum money) is stored separately, which has no direct relationship
with the quantum registers. On the contrary, in our protocol the classical
data is stored directly in the quantum registers and is perfectly readable
by any user. Thus it is more suitable for anti-piracy protection of
classical data.

\end{document}